\documentclass[12pt]{article}

\usepackage{newtxtext,newtxmath}

\usepackage{graphicx}

\usepackage[letterpaper,margin=1in]{geometry}

\usepackage{bm}

\renewenvironment{abstract}
	{\quotation}
	{\endquotation}

\date{}

\makeatletter
\renewcommand{\fnum@figure}{\textbf{Figure \thefigure}}
\renewcommand{\fnum@table}{\textbf{Table \thetable}}

\newcommand{\CsVSb}{{CsV$_{3}$Sb$_{5}$}}

\newcommand{\RbVSb}{{RbV$_{3}$Sb$_{5}$}}
\newcommand{\AVSb}{{AV$_{3}$Sb$_{5}$}} 
\newcommand{\parallelsum}{\mathbin{\!/\mkern-5mu/\!}}

\makeatother

\usepackage{scicite}

\usepackage{url}

\def\scititle{
	Thickness-driven crossover from conventional to chiral nonreciprocal superconductivity in kagome metal CsV$_{3}$Sb$_{5}$
}
\title{\bfseries \boldmath \scititle}

\author{
	Wei~Zhang$^{1\dagger\ast}$,
Jiangbo~Luo$^{1,2\dagger}$,
Nikolai~Peshcherenko$^3$,
Zheyu~Wang$^4$\and
Chun~Wai~Tsang$^4$,
Kwing~To~Lai$^4$,
King~Yau~Yip$^1$,
Kenji~Watanabe$^6$\and
Takashi~Taniguchi$^7$,
Junxiong~Hu$^1$,
Yang~Zhang$^1$,
Swee~K.~Goh$^{4,5}$,
A.~Ariando$^{1\ast}$\and
	\small$^{1}$Department of Physics, Faculty of Science, National University of Singapore, 117542, Singapore.\and
\small$^{2}$State Key Laboratory of Chemo and Biosensing, College of Chemistry and Chemical Engineering,\and 
\small Hunan University, Changsha 410082, China.\and
\small$^{3}$Max Planck Institute for Chemical Physics of Solids, Dresden 01187, Germany.\and
\small$^{4}$Department of Physics, The Chinese University of Hong Kong, Shatin, Hong Kong 999077, China.\and
\small$^{5}$State Key Laboratory of Quantum Information Technologies and Materials,\and 
\small The Chinese University of Hong Kong, Shatin, Hong Kong 999077, China.\and
\small$^{6}$Research Center for Electronic and Optical Materials,\and
\small National Institute for Materials Science, Tsukuba 305-0044, Japan.\and
\small$^{7}$Research Center for Materials Nanoarchitectonics,\and
\small National Institute for Materials Science, Tsukuba 305-0044, Japan.\and
\small$^\ast$Corresponding author. Email: wzhang@nus.edu.sg; ariando@nus.edu.sg\and
\small$^\dagger$These authors contributed equally to this work.
}

\begin{document} 

\maketitle

\newpage
\begin{abstract} \bfseries \boldmath
Superconductivity and its potential applications are governed by the symmetry of the superconducting order parameter. In the kagome metal \CsVSb, most bulk studies indicate conventional $s$‑wave pairing. However, ultrathin flakes exhibit nonreciprocal transport, in particular a zero-field superconducting diode effect, which requires broken inversion and time-reversal symmetries. Here, using thickness dependent transport measurements, we observe the emergence of nonreciprocal second‑harmonic magnetotransport signals and a zero‑field superconducting diode effect, accompanied by a pronounced reduction of the out-of-plane coherence length with decreasing thickness. Upper critical field measurements further reveal a dimensional crossover from three‑dimensional superconductivity in bulk to two‑dimensional superconductivity in thin flakes. These findings indicate a thickness-induced chiral superconducting phase that breaks both inversion and time-reversal symmetries in the two‑dimensional limit. Our work not only clarifies long-standing controversies regarding the pairing symmetry in \CsVSb, but also establishes thin‑flake kagome superconductors as a versatile platform for engineering nonreciprocal quantum devices and exploring emergent topological phases.
\end{abstract}

\noindent

The kagome lattice, composed of corner-sharing triangles, naturally hosts flat bands, van Hove singularities, and Dirac points, providing a fertile platform for exotic quantum phenomena~\cite{Zhou2017, Kiesel2012, Kiesel2013, Wang2013}. The recently discovered kagome superconductors \AVSb\ (A = K, Rb, Cs) have attracted significant attention, where vanadium atoms form a perfect kagome net (Fig.~\ref{fig1}A). Among the three compounds of \AVSb, \CsVSb\ exhibits the highest superconducting transition temperature ($T_{\rm c}\sim$3 K) at ambient pressure~\cite{Ortiz2019, Ortiz2020, Chen2021a, Li2021, Liang2021, Zhao2021, Zhou2021, Wang2025, Kang2022, Wang2024, Yu2021, Zhang2024, Yin2022, Wilson2024, Wang2025, Xu2025}. 

In \CsVSb, a charge-density-wave (CDW) order emerges below $\sim$90~K. Early measurements including muon spin relaxation, the anomalous Hall effect, and the magneto-optical Kerr effect, suggested time-reversal symmetry breaking (TRSB) within the CDW phase, leading to the proposed chiral flux phase with loop currents~\cite{Yu2021c, Yu2021b, Xu2022, Wu2022, Feng2021}. However, a recent high-resolution magneto-optical Kerr effect study failed to detect TRSB signals~\cite{Saykin2023}. Concurrently, detailed transport studies attributed the anomalous signals to multi-band effects originating from a small, high-mobility Fermi pocket~\cite{Liu2025}. Consequently, whether TRSB occurs in the CDW phase remains intensely debated.

The superconducting (SC) pairing symmetry in \CsVSb\ is also controversial. Nuclear quadrupole resonance and nuclear magnetic resonance measurements show a decrease of the Knight shift and a Hebel–Slichter coherence peak below $T_{\rm c}$~\cite{Mu2021}. The insensitivity of $T_{\rm c}$ to non-magnetic impurities, along with tunneling diode oscillator measurements, self-field critical current observations, and scanning tunneling microscopy results collectively point to conventional multi-gap $s$-wave superconductivity~\cite{Gupta2022, Gupta2022a, Duan2021, Roppongi2023, Xu2021, Zhang2023}. In contrast, the observation of nonreciprocal signals and the zero-field superconducting diode effect (SDE) indicate inversion ($\mathcal{P}$) and time-reversal ($\mathcal{T}$) symmetry breaking in the SC state~\cite{Wu2022, Le2024}. Superconducting interference patterns under magnetic fields and the intrinsic Josephson effect have also been detected in \CsVSb~\cite{Le2024, Ge2025, Lou2026}. Both the SDE and interference patterns can be modulated by thermal cycling, suggesting dynamic superconducting domains. Together, these results strongly imply an unconventional superconductivity in \CsVSb.

Notably, most studies reporting unconventional SC properties were conducted on ultra-thin \CsVSb\ flakes~\cite{Le2024, Ge2025, Lou2026, Wu2022, Hossain2025}. Therefore, a systematic investigation of thickness-dependent SC properties is essential. In this work, we study superconductivity in CsV$_3$Sb$_5$ across a range of thicknesses. Our second-harmonic transport measurements reveal the emergence of nonreciprocity within the SC transition region in thin flakes below a critical thickness of $\sim$100 nm. $V\mbox{-}I$ measurements further show the emergence of a zero$\mbox{-}$field SDE below $\sim$100 nm. Additionally, angular-dependent upper critical field ($H_{\rm c2}$) results demonstrate that while bulk \CsVSb\ exhibits three-dimensional superconductivity, thin flakes display two-dimensional characteristics, with the SC anisotropy factor increasing as thickness decreases. Surprisingly, the out-of-plane coherence length ($\xi_{\rm c}$) decreases significantly with the reduction of thickness and becomes comparable to the out-of-plane lattice constant for the ultrathin flakes. These findings indicate the emergence of a thickness-induced chiral superconducting phase that breaks both $\mathcal{P}$ and $\mathcal{T}$ symmetries in the two-dimensional limit. Our work resolves long-standing ambiguities regarding the pairing symmetry of \CsVSb\ and establishes thin-flake kagome metals as a highly tunable platform.

\subsection*{Nonreciprocal second-harmonic transport}

We performed second-harmonic measurements on \CsVSb\ flakes using a Stanford Research 830 lock-in amplifier. Figures~\ref{fig1}B--E summarize the results for Sample 4 (S4), a 36 nm-thick \CsVSb~flake. The sample exhibits a SC transition with $T_{\rm c,onset}=4.5$~K (Fig.~\ref{fig1}B), consistent with values reported in the literature~\cite{Sun2025, Song2023}. Figure~\ref{fig1}C shows the raw second harmonic data, $V_{xx}^{2\omega}$, at 1.8~K with the magnetic field parallel to the kagome plane. As indicated by the arrows, pronounced peak-like features appear at positive magnetic field; upon reversing the field, these peaks become valley-like, demonstrating the nonreciprocal nature of the second-harmonic response. Similar nonreciprocal signals are also observed for $H\parallelsum c$, as shown in Fig.~\ref{fig1}D. The nonreciprocal response weakens and eventually disappears as the SC state is suppressed by increasing temperature, as summarized by the temperature dependence of the nonlinear coefficient $\gamma$ for the most pronounced peak (Fig.~\ref{fig1}E). This behaviour indicates that the nonreciprocity is associated with superconductivity in ultrathin flakes.

To investigate how thickness influences nonreciprocity, we systematically measured the second-harmonic magnetotransport response of \CsVSb\ flakes with different thicknesses (Fig.~\ref{fig2}). For the 36~nm flake, the nonreciprocal response occurs within the SC transition region, as indicated by comparison with the field-dependent resistance $R_{xx}(H)$ in Fig.~\ref{fig2}, M to P. With increasing thickness, nonreciprocal signals remain detectable in the 98~nm flake within the SC transition region (Fig.~\ref{fig2}, I to L), whereas no nonreciprocal response is observed in the 240~nm (Fig.~\ref{fig2}, F and G) and 435~nm (Fig.~\ref{fig2}, B and C) flakes.

To quantify the nonreciprocity, we extracted the odd component of the raw second-harmonic signal, a procedure commonly adopted in analyses of nonreciprocal transport (e.g., ion-gated MoS$_2$)~\cite{Wakatsuki2017}. Figures~\ref{fig2}, N and O show the odd components of $V_{xx}^{2\omega}$ at different temperatures in S4 with $H\parallelsum ab$ and $H\parallelsum c$, respectively. Several pairs of sign-reversal features are resolved within the SC transition region. This behaviour differs from ion-gated MoS$_2$, where only one sign-reversal pair was reported over a range of temperatures. In addition, Fig.~\ref{fig1}E together with Fig.~\ref{fig2}, M to P shows that the nonreciprocal signals weaken and disappear as the SC state is suppressed by increasing temperature, further supporting their superconducting origin.

A constant current of 100~$\mu$A was applied to all samples, resulting in a lower current density and reduced signal-to-noise ratio for thicker flakes. However, the signal magnitude observed in the thinner samples indicates that this effect alone cannot account for the absence of nonreciprocity in the 240~nm and 435~nm samples. This conclusion is further supported by measurements at higher current density (fig.~\ref{FigS1}), which significantly improve the signal-to-noise ratio but still fail to reveal nonreciprocal signals in the 435~nm flake. Because nonreciprocal transport requires both inversion-symmetry breaking and time-reversal-symmetry breaking, our second-harmonic results suggest at least inversion symmetry breaking in the SC state below a critical thickness of $\sim 100$~nm, while time-reversal symmetry may be broken either intrinsically or by the applied magnetic field.

\subsection*{Zero-field superconducting diode effect}

To test for possible spontaneous time-reversal symmetry breaking in \CsVSb, we probed the zero-field SDE. Because our second-harmonic measurements indicate that nonreciprocal superconductivity emerges only below $\sim 100$~nm, we examined the SDE in \CsVSb\ flakes with different thicknesses. 

Before the $V$--$I$ measurements, the superconducting magnet was warmed to room temperature to eliminate any residual magnetic field. We first measured a 155~nm-thick flake under zero applied field. A Keithley 6221 current source operated in pulsed delta mode was used to minimize Joule heating. As shown in Fig.~\ref{fig3}A, the voltage remains near zero as the current increases from zero; once the current exceeds a threshold, the voltage rises sharply, indicating a transition from the superconducting to the normal state. To check for heating-related artifacts, we measured the $V$--$I$ curves using both $I_- \rightarrow 0\rightarrow I_+$ and $I_+\rightarrow 0 \rightarrow I_-$ current sequences. The two traces overlap closely, indicating negligible heating.

We then compare the $0 \rightarrow I_{+}$ and $0 \rightarrow I_{-}$ branches. As shown in Fig.~\ref{fig3}E, these branches overlap, indicating the absence of an SDE in the 155~nm flake. By contrast, as the thickness is reduced, a small but reproducible mismatch between the $0 \rightarrow I_{+}$ and $0 \rightarrow I_{-}$ branches appears in the 100~nm flake below 3.5~K (Fig.~\ref{fig3}F). A clear SDE is observed in even thinner samples (Fig.~\ref{fig3}, G and H). The SDE in the 65~nm and 30~nm flakes is also evident in the ${\rm d}V/{\rm d}I$ versus $I$ curves, where it manifests as a characteristic tilted ``8'' shape (Fig.~\ref{fig3}, K and L). Together, these $V$--$I$ measurements demonstrate the emergence of a zero-field SDE with decreasing thickness, implying that both inversion symmetry and time-reversal symmetry are broken in the superconducting state for flakes thinner than $\sim 100$~nm.

\subsection*{Angular dependence of upper critical field}

Our second-harmonic and $V$--$I$ measurements suggest that thin \CsVSb\ flakes below $\sim 100$~nm host a SC state with broken inversion and time-reversal symmetries, distinct from the conventional $s$-wave superconductivity reported in the bulk~\cite{Gupta2022, Gupta2022a, Duan2021, Roppongi2023, Xu2021, Zhang2023}. To further characterise the thickness-dependent SC properties of \CsVSb, we studied the angular dependence of the upper critical field $H_{\rm c2}$ in bulk crystals and thin flakes. 

All measurements were carried out at temperatures corresponding to a reduced temperature of $T/T_{\rm c,onset}\sim 0.4$ (1.4~K for the bulk crystal, and 1.8~K or 1.9~K for the flakes). The temperature dependence of the resistance for the bulk and flakes is shown in fig.~\ref{FigS2}, and representative field-dependent resistance curves at selected angles are shown in fig.~\ref{FigS3}. We determine $H_{\rm c2}$ from the zero-resistance criterion, taking the average of the values obtained for positive and negative field polarities.

Figure~\ref{fig4}A shows $H_{\rm c2}(\theta)$ for bulk \CsVSb, where $\theta$ is the angle between the magnetic field and the kagome plane (so that $\theta=0^\circ$ corresponds to $H\parallelsum ab$ and $\theta=90^\circ$ to $H\parallelsum c$). For $H\parallelsum c$ ($\theta=90^\circ$), $H_{\rm c2}$ is $\sim 0.2$~T, consistent with previous reports~\cite{Fukushima2024, Ni_2021}. As the field is tilted away from the $c$ axis toward the $ab$ plane, $H_{\rm c2}$ increases monotonically and reaches a maximum for $H\parallelsum ab$ ($\theta=0^\circ$), reflecting the layered structure. The anisotropy factor, defined as $\Gamma=H_{\rm c2}(0^\circ)/H_{\rm c2}(90^\circ)$, is 6.9, in agreement with earlier bulk studies~\cite{Fukushima2024, Ni_2021}. 

The angular dependence of $H_{\rm c2}$ is commonly described by either the two-dimensional (2D) Tinkham model or the three-dimensional (3D) anisotropic-mass Ginzburg--Landau (GL) model~\cite{Tinkham2004, Goh2012}:
\begin{equation*}
\left[\frac{H_{\rm c2}(\theta)\cos\theta}{H_{\rm c2}(0^\circ)}\right]^2=1-\alpha\left|\frac{H_{\rm c2}(\theta)\sin\theta}{H_{\rm c2}(90^\circ)}\right|-\beta\left[\frac{H_{\rm c2}(\theta)\sin\theta}{H_{\rm c2}(90^\circ)}\right]^2
\end{equation*}
where $(\alpha,\beta)=(1,0)$ corresponds to the Tinkham model and $(\alpha,\beta)=(0,1)$ to the 3D anisotropic-mass GL model. The Tinkham model predicts a cusp at $\theta=0^\circ$, whereas the 3D GL model yields a smooth angular dependence. For the bulk crystal (Fig.~\ref{fig4}A), the data deviate strongly from the Tinkham fit but are well described by the 3D GL model, indicating intrinsically three-dimensional superconductivity.

We next examine thin flakes. For the 435~nm flake (Fig.~\ref{fig4}B), $H_{\rm c2}(\theta)$ exhibits a clear cusp around $\theta=0^\circ$ and is well described by the Tinkham model. The anisotropy factor increases to $\Gamma=12.1$, about 1.8 times the bulk value, indicating a crossover to 2D superconductivity. This result is unexpected because $H_{\rm c2}(0^\circ)=3.02$~T and $H_{\rm c2}(90^\circ)=0.23$~T correspond to coherence lengths of $\xi_{\rm ab}=37.5$~nm and $\xi_{\rm c}=2.9$~nm, respectively, and the flake thickness (435~nm) is far larger than $\xi_{\rm c}$. Notably, $\xi_{\rm c}$ is only about three times the out-of-plane lattice constant ($c=9.3$~\AA), which is consistent with quasi-2D superconductivity. In addition, a recent study on \RbVSb\ suggests that the SC dimensionality can be highly sensitive to biaxial substrate-induced strain~\cite{Poon2025}. Because our thin \CsVSb\ flakes are supported on silicon substrates, biaxial strain may contribute to the observed dimensional crossover by tuning the interlayer coupling.

We further investigated $H_{\rm c2}(\theta)$ across multiple thicknesses. As shown in Fig.~\ref{fig4}, B to E, all thin flakes are better described by the 2D Tinkham model. For the 37~nm flake (Fig.~\ref{fig4}E), however, the difference between the 2D and 3D fits becomes difficult to resolve within our experimental uncertainty, which is common in layered superconductors with large anisotropy~\cite{Naughton1988}. 

We also measured the temperature dependence of $H_{\rm c2}$ for $H\parallelsum ab$ and $H\parallelsum c$. As shown in Fig.~\ref{fig4}F, the bulk crystal exhibits an upturn in $H_{\rm c2}(T)$ for $H\parallelsum ab$ (arrow), a feature previously interpreted as evidence for multiband superconductivity in \CsVSb~\cite{Ni_2021}. This upturn persists in the 435~nm and 240~nm flakes but weakens in thinner flakes (98~nm and 37~nm), where $H_{\rm c2}(T)$ becomes nearly linear below $T_{\rm c}$. Using these $H_{\rm c2}(T)$ data, we extract the temperature dependence of $\Gamma$. Figure~\ref{fig4}M summarises $\Gamma$ as a function of thickness at $T/T_{\rm c,onset}\sim 0.4$, where $\Gamma$ has already saturated (fig.~\ref{FigS4}). Notably, $\Gamma$ increases systematically as the thickness decreases; the thinnest sample (S5, 37~nm) reaches $\Gamma=17.5$, about 2.5 times the bulk value.

We note that in-plane anisotropy could, in principle, influence the extracted $\Gamma$, because $H_{\rm c2}$ may depend on the in-plane field orientation. However, recent magnetotransport studies show that the in-plane $H_{\rm c2}$ in \CsVSb\ is mainly governed by the relative angle between the in-plane field and the current direction~\cite{Yao2025}. In our measurements, the magnetic field was always applied perpendicular to the current for all samples. Therefore, the observed enhancement of $\Gamma$ with decreasing thickness reflects an intrinsic evolution of superconductivity in \CsVSb.

\subsection*{Discussion and Conclusion}

Our observation of (i) nonreciprocal second-harmonic signals, (ii) a zero-field superconducting diode effect, and (iii) a systematic increase in the anisotropy factor $\Gamma$ collectively points to a thickness-driven dimensional crossover in \CsVSb. Specifically, the data suggest an evolution from a conventional, three-dimensional, fully gapped $s$-wave state in the bulk~\cite{Mu2021, Gupta2022, Gupta2022a, Duan2021, Roppongi2023, Xu2021, Zhang2023} to a two-dimensional superconducting state with broken inversion symmetry ($\mathcal{P}$) and time-reversal symmetry ($\mathcal{T}$) in flakes thinner than $\sim 100$~nm. 

\begin{figure}[!t]\centering
      \resizebox{14cm}{!}{
              \includegraphics{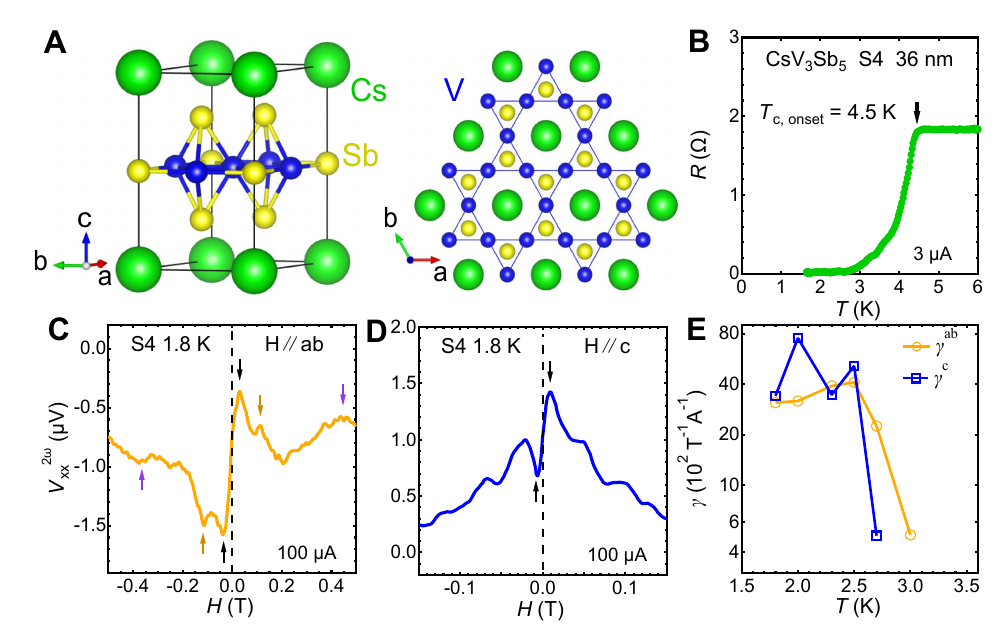}}                				
              \caption{\label{fig1}  
        \textbf{Nonreciprocal second-harmonic signals in a 36 nm \CsVSb\ flake.} (\textbf{A}) Crystal structure of \CsVSb. (\textbf{B}) Temperature dependence of resistance for a \CsVSb\ (S4) thin flake with $T_{\rm c, onset}$ of 4.5 K. (\textbf{C} and \textbf{D}) Raw data of $V_{xx}^{2\omega}$ detected by lock-in amplifier at 1.8~K with $H\parallelsum ab$ (C) and $H\parallelsum c$ (D). The arrows indicate the nonreciprocal peaks or valleys. (\textbf{E}) Temperature dependence of the nonlinear coefficient $\gamma$,  for the most pronounced peak (black arrows) in $H\parallelsum ab$ and $H\parallelsum c$ direction. }
\end{figure}

\begin{figure*}[!t]\centering
      \resizebox{16.5 cm}{!}{
 \includegraphics{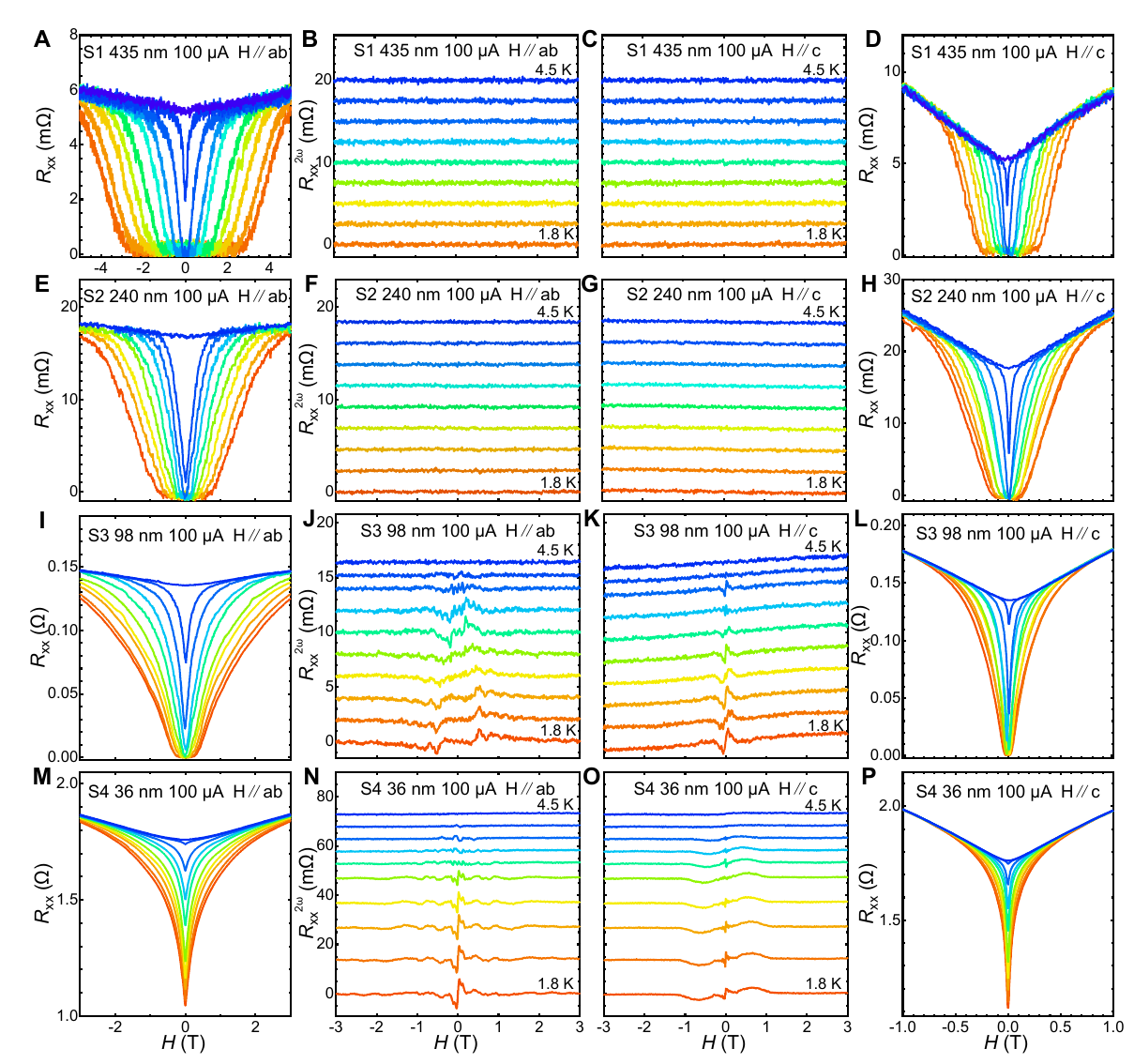}}       
 \caption{\label{fig2} 
\textbf{Thickness dependence of nonreciprocal signals.} Resistance and the odd components of the second harmonic signals in \CsVSb\ thin flakes with the thickness of 435 nm (\textbf{A} to \textbf{D}), 240 nm (\textbf{E} to \textbf{H}), 98 nm (\textbf{I} to \textbf{L}) and 36 nm (\textbf{M} to \textbf{P}).}            
\end{figure*}

To elucidate the microscopic origin of the observed nonreciprocity, it is useful to distinguish between the finite-field and zero-field regimes. In the presence of an in-plane magnetic field (Figs.~\ref{fig1} and~\ref{fig2}), $\mathcal{T}$ symmetry is explicitly broken. In noncentrosymmetric superconductors, the interplay between Zeeman coupling and spin--orbit interaction can induce a finite centre-of-mass momentum $\bm{q}\neq 0$ for Cooper pairs, leading to a helical superconducting state~\cite{SDE_Liang_Fu,nagaosa_SDE_Rashba}. Such finite-momentum pairing manifests as a directional dependence of the critical current and resistance, giving rise to the second-harmonic magnetotransport response observed here (see Materials and Methods). Related layer-dependent finite-$\bm{q}$ superconducting states have also been proposed for misfit-layer superconductors~\cite{bilayer}, highlighting that the quasi-two-dimensional nature of thin \CsVSb\ flakes plays a crucial role in stabilizing this state.

\begin{figure*}[!t]\centering
       \resizebox{16.5 cm}{!}{
              \includegraphics{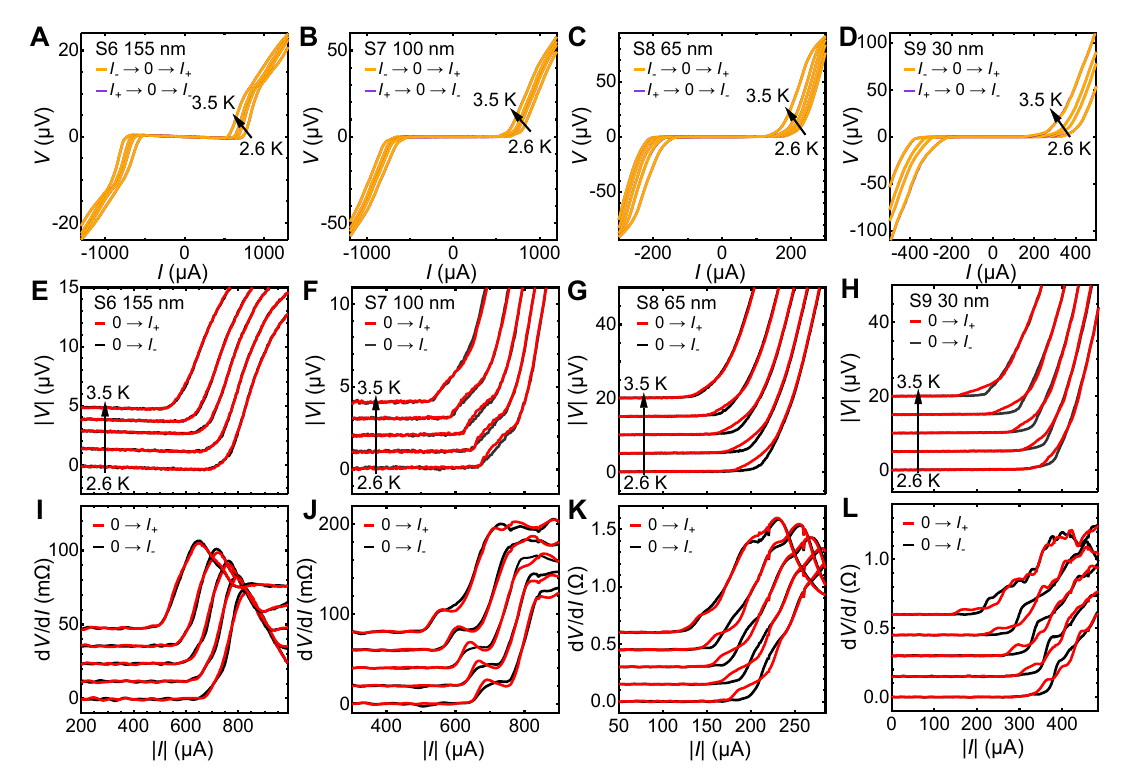}}                			
              \caption{\label{fig3}
\textbf{Zero-field superconducting diode effect.} (\textbf{A} to \textbf{D}) Zero$\mbox{-}$field $V\mbox{-}I$ curves measured in $I_- \rightarrow 0\rightarrow I_+$ and $I_+\rightarrow 0 \rightarrow I_-$ sequences for 155 nm (A), 100 nm (B), 65 nm (C) and 30 nm (D) thin flakes at various temperatures. (\textbf{E} to \textbf{H}) The $0\rightarrow I_+$ and $0 \rightarrow I_-$ branches of 155 nm (E), 100 nm (F), 65 nm (G) and 30 nm (H) thin flakes, respectively. (\textbf{I} to \textbf{L}) The calculated first derivative of $V(I)$ for the 155 nm (I), 100 nm (J), 65 nm (K) and 30 nm (L) thin flakes, respectively.}
\end{figure*}

\begin{figure*}[!t]\centering
       \resizebox{16.6 cm}{!}{
              \includegraphics{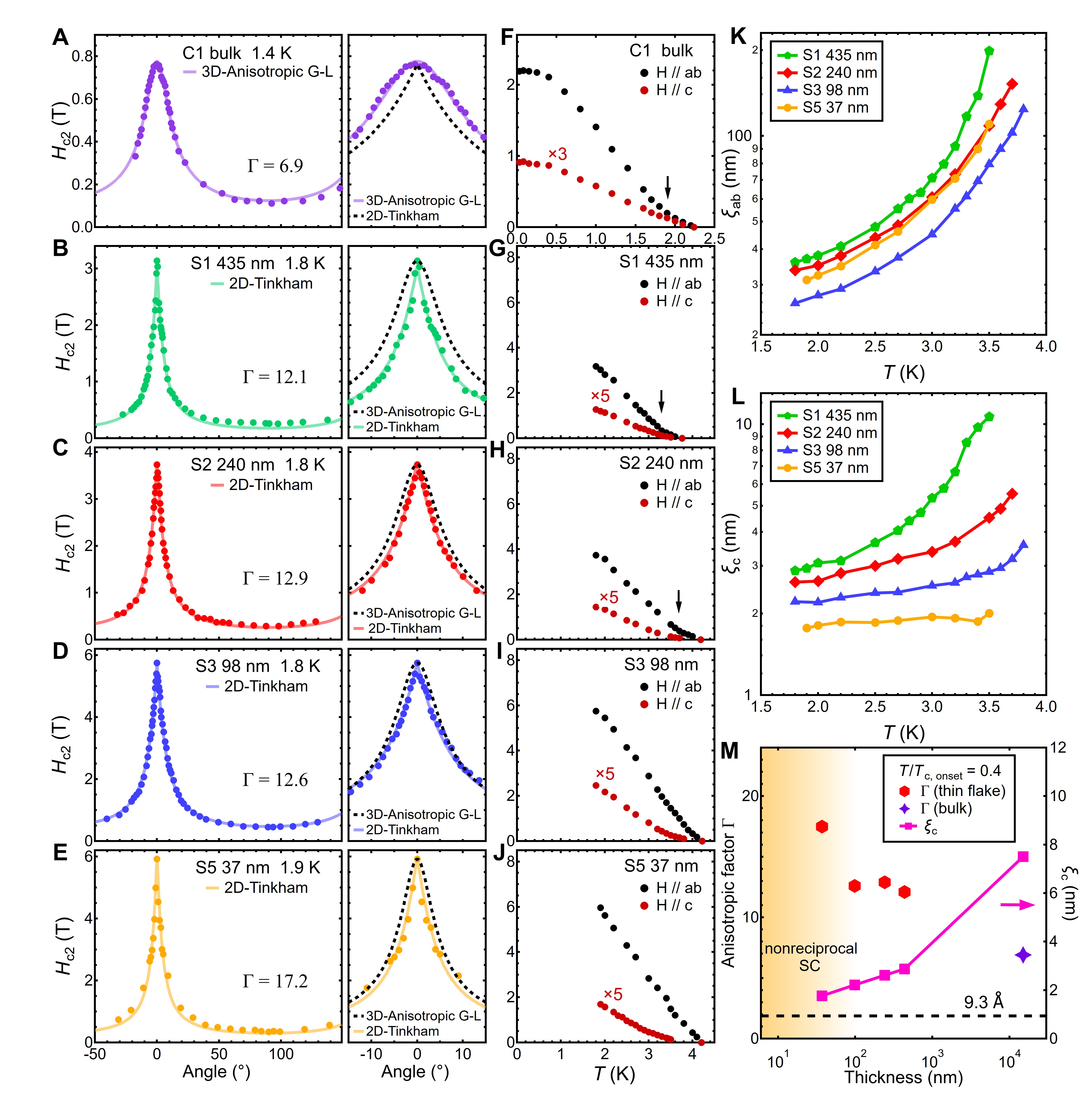}}                			
              \caption{\label{fig4}
\textbf{Upper critical field measurements.} (\textbf{A} to \textbf{E}) Angular dependence of $H_{c2}$ in bulk (A) and thin flakes (B to E). The dashed lines represent simulation results to the respective models. (\textbf{F} to \textbf{J}) Temperature dependence of in-plane and out-of-plane $H_{\rm c2}$ for bulk (F) and thin flakes (G to J).  (\textbf{K} and \textbf{L}) Temperature dependence of the coherence length $\xi_{\rm ab}$ (K) and $\xi_{\rm c}$ (L). (\textbf{M}) Thickness dependence of the anisotropic factor $\Gamma$ and the coherence length $\xi_{\rm c}$ at the reduced temperature of 0.4. The dashed line indicates the out-of-plane lattice constant of \CsVSb.}
\end{figure*}

The observation of the superconducting diode effect at zero external field (Fig.~\ref{fig3}) imposes a stricter constraint, namely spontaneous breaking of $\mathcal{T}$ symmetry. Although bulk \CsVSb\ is widely regarded as a fully gapped $s$-wave superconductor~\cite{Mu2021, Gupta2022, Gupta2022a, Duan2021, Roppongi2023, Xu2021, Zhang2023}, our results indicate a qualitative departure from this behaviour in the two-dimensional limit. In ultra-thin flakes, the reduced interlayer coupling can strongly modify how superconducting order parameters on different layers lock to one another. Consistent with this picture, Figs.~\ref{fig4}, K to M show that while the in-plane coherence length $\xi_{\rm ab}$ does not exhibit a clear systematic trend, the out-of-plane coherence length $\xi_{\rm c}$ decreases markedly with decreasing thickness. For the 37~nm flake, $\xi_{\rm c}\approx 1.8$~nm, approaching the out-of-plane lattice constant ($c=9.3$~\AA). In this regime, interlayer phase stiffness is reduced and phase fluctuations are enhanced, allowing the superconducting order parameters on different layers to adopt nontrivial relative phases. 

Our phenomenological free-energy analysis (Materials and Methods) indicates that, in the thin limit, certain interlayer interaction potentials can favour a spontaneous phase difference between layers. This interlayer frustration can stabilise a chiral superconducting state (for example, an $s+is$ state emerging from the bulk $s$-wave parent state) that intrinsically breaks $\mathcal{T}$. Such a state naturally accounts for the nonvanishing nonreciprocity and the superconducting diode effect observed even in the absence of an external magnetic field.

By tracing the evolution of the superconducting state across different sample thicknesses, our study resolves long-standing debates regarding the pairing symmetry of \CsVSb. We show that the nonreciprocal transport in thin flakes is not an extrinsic anomaly but a signature of an emergent chiral phase that is fundamentally distinct from the bulk $s$-wave state. More broadly, this dimensional crossover underscores the critical role of thickness in kagome metals and establishes thin \CsVSb\ as a versatile platform for on-chip nonreciprocal quantum devices and for exploring chiral and topological superconducting phases.
\newpage

\clearpage 

\providecommand{\noopsort}[1]{}\providecommand{\singleletter}[1]{#1}%


\section*{Acknowledgments}
\paragraph*{Funding:}
This work was supported by the Ministry of Education (MOE), Singapore, under its Tier-2 Academic Research Fund (AcRF), Grants No. MOE-T2EP50123-0013, and by the National Research Foundation (NRF) Investigatorship Award on "Quantum design of superconductivity and correlated phases in layered complex oxides". The work at The Chinese University of Hong Kong was supported by Research Grants Council of Hong Kong (14302724 and 14301725). N.P. and Y.Z. are supported by Max Planck partner lab for quantum materials. K.W. and T.T. acknowledge support from the JSPS KAKENHI (Grant Numbers 21H05233 and 23H02052), the CREST (JPMJCR24A5), JST and World Premier International Research Center Initiative (WPI), MEXT, Japan.
\paragraph*{Author contributions:}
W.Z., S.K.G. and A.A. conceived the project. W.Z., J.L., Z.W. conducted the experiments and analyzed the data. C.W.T. and K.T.L. prepared \CsVSb\ samples. T.T. and K.W. provided the bulk hBN crystals. N.P. and Y.Z. provided theoretical support. W.Z., Y.Z. and A.A. wrote the manuscript with input from all authors.
\paragraph*{Competing interests:}
The authors declare no competing interests.
\paragraph*{Data and materials availability:}
The data that support the findings of this study are available from the corresponding author upon reasonable request.


\subsection*{Supplementary materials}
Materials and Methods\\
Figures. S1 to S5\\


\newpage


\renewcommand{\thefigure}{S\arabic{figure}}
\renewcommand{\thetable}{S\arabic{table}}
\renewcommand{\theequation}{S\arabic{equation}}
\renewcommand{\thepage}{S\arabic{page}}
\setcounter{figure}{0}
\setcounter{table}{0}
\setcounter{equation}{0}
\setcounter{page}{1} 


\begin{center}
\section*{Supplementary Materials for\\ \scititle}

Wei~Zhang$^{1\dagger\ast}$,
Jiangbo~Luo$^{1,2\dagger}$,
Nikolai~Peshcherenko$^3$,
Zheyu~Wang$^4$\\
Chun~Wai~Tsang$^4$,
Kwing~To~Lai$^4$,
King~Yau~Yip$^1$,
Kenji~Watanabe$^6$\\
Takashi~Taniguchi$^7$,
Junxiong~Hu$^1$,
Yang~Zhang$^1$,
Swee~K.~Goh$^{4,5}$,
A.~Ariando$^{1\ast}$\\
\small$^\ast$Corresponding author. Email: wzhang@nus.edu.sg; ariando@nus.edu.sg\\
\small$^\dagger$These authors contributed equally to this work.

\end{center}

\subsubsection*{This PDF file includes:}
Materials and Methods\\
Figures S1 to S5\\


\newpage


\subsection*{Materials and Methods}


\subsubsection*{Crystal growth and device fabrication}

High-quality single crystals of \CsVSb~were synthesized from high-purity Cs (99.95 \% ingot), V (99.9 \% powder) and Sb (99.9999 \% shot) using the self-flux method~\cite{Ortiz2019,Ortiz2020}. The raw materials were sealed inside a pure-Ar-filled stainless steel jacket with the molar ratio of Cs:V:Sb = 7:3:14. The flakes were exfoliated from bulk \CsVSb\ crystals and transferred to pre-patterned  electrodes on Si/SiO$_2$ substrates. A layer of hexagonal boron nitride (h-BN) was transferred on top for encapsulation. The thickness of the flake was determined by atomic force microscopy (fig.~\ref{FigS5}).

\subsubsection*{Transport measurements}

Magnetotransport properties of thin flakes were measured in an Oxford TeslatronPT cryostat. Bulk sample measurements were performed in a Bluefors dilution refrigerator. A Stanford Research 830 lock-in amplifier with frequency of 113 Hz was used for the second-harmonic measurements. For $V\mbox{-}I$ curves, a Keithley 6221 current source (pulsed delta mode, 10 ms pulse width, 1 s repetition time) and a Keithley 2182A nanovoltmeter were used to minimize Joule heating. For zero-field SDE measurements, the magnet was warmed to room temperature to eliminate residual fields.

\subsubsection*{Time-reversal symmetry breaking}

From a symmetry perspective, the observation of non-reciprocal supercurrent requires the breaking of both time-reversal and inversion symmetries. While the inversion symmetry breaking could be induced by the interface, the origin of time-reversal symmetry breaking deserves further discussion. In the following two sections, we consider two scenarios for the time-reversal symmetry breaking: one induced by an applied magnetic field and the other from intrinsic time-reversal symmetry breaking state in a bilayer system with the $s+is$ superconductivity.

\noindent \textbf{Nonreciprocal transport phenomena due to an external magnetic field.} 
Based on the evidence of non-reciprocal second harmonic behavior in the presence of magnetic field (as shown in Fig.~\ref{fig1} and Fig.~\ref{fig2}), we suggest that the applied magnetic field leads to finite momentum superconductivity, allowing for non-reciprocal supercurrents. Given the layered structure of \CsVSb, the 2D superconductivity observed in the thin flakes, the short out-of-plane coherence length ($\xi_{\rm c}$) in our ultra-thin \CsVSb\ thin flake (Fig.~\ref{fig4}), comparable to the out-of-plane lattice constant  and the fact that non-reciprocal effects only appear in flakes thinner than $\sim$100 nm, we present a theory for two-dimensional superconductors.

Following previous works \cite{nagaosa_SDE_Rashba,SDE_Liang_Fu}, one can write down a minimal Ginzburg-Landau free energy that would allow for superconducting diode effect:
\begin{align}
    \mathcal{F}=\int\frac{d^2\mathbf{q}}{(2\pi)^2}\mathcal{F}_\mathbf{q}=\int\frac{d^2\mathbf{q}}{(2\pi)^2}\left(\sigma(\mathbf{q})|\Delta_\mathbf{q}|^2+\frac{1}{2}\gamma|\Delta_\mathbf{q}|^4\right)
    \label{eq:GL_inplane},
\end{align}
with the coefficient $\sigma(\mathbf{q})$ depending on magnetic field $\mathbf{H}$:
\begin{align}
    \sigma(\mathbf{q})=a+b(\mathbf{n}\times\mathbf{H})\cdot\mathbf{q},\quad b=b_0+b_1\mathbf{q}^2,\quad a=t+a_0\mathbf{q}^2,\quad t=\frac{T-T_c}{T_c}<0,
\end{align}
where $\mathbf{n}$ represents normal vector to the two-dimensional plane. The Ginzburg-Landau functional from Eq. \eqref{eq:GL_inplane} predicts finite momentum superconductivity at momentum $\mathbf{q}_0$:
\begin{align}
    \mathbf{q}_0=-\frac{b}{2a_0}(\mathbf{n}\times\mathbf{H}).
\end{align}

Within this minimalistic model one can then show that the results for supercurrent $\mathbf{J}(\mathbf{q})=2e\partial_\mathbf{q}\mathcal{F}_\mathbf{q}$, allows for the efficiency $\eta=\frac{I_c^+-I_c^-}{I_c^++I_c^-}$ of superconducting diode effect \cite{SDE_Liang_Fu}:
\begin{align}
    \eta=\sqrt{\frac{|t|}{3a_0}}\frac{b_1}{a_0}(\mathbf{n}\times\mathbf{H})\cdot\mathbf{i},
\end{align}
with $\mathbf{i}$ being the current direction.\\

\noindent \textbf{Layer-dependent superconducting order parameter.}
We now turn to magnetic field-unrelated time reversal symmetry breaking linked to a two-component superconducting state. Namely, due to the layered structure of the material and the short out-of-plane coherence length ($\xi_{\rm c}$) in our ultra-thin \CsVSb\ thin flake (Fig.~\ref{fig4}), the superconducting order parameter could be in principle layer-dependent. The corresponding minimal Ginzburg-Landau free energy model then reads
\begin{align}
    F=F_1+F_2+F_\mathrm{int},\quad F_1=\alpha_1|\eta_1|^2+\frac{\beta_1}{2}|\eta_1|^4,\quad F_2=\alpha_2|\eta_2|^2+\frac{\beta_2}{2}|\eta_2|^4, \nonumber\\
    F_\mathrm{int}=\gamma(\eta_1^{2}\eta_2^{*2}+\eta_1^{*2}\eta_2^{2})=2\gamma |\eta_1|^2|\eta_2|^2\cos(2(\varphi_1-\varphi_2)),
    \label{eq:GL_phases}
\end{align}
where $\varphi_{1,2}$ are phases of the corresponding superconductor order parameters:
\begin{align}
    \eta_{1,2}=|\eta_{1,2}|e^{i\varphi_{1,2}}
\end{align}
The Ginzburg-Landau functional from \eqref{eq:GL_phases} for $\gamma>0$ reaches its minimum at the finite phase difference
\begin{align}
    \varphi_1-\varphi_2=\frac{\pi}{2},
\end{align}
suggesting for $s+is$ two-component superconductivity in a bilayer system. This state, in turn, allows for a spontaneous time-reversal symmetry breaking and the observed non-reciprocity.




\newpage

\begin{figure}[h]\centering
              \includegraphics[width=1.02\textwidth]{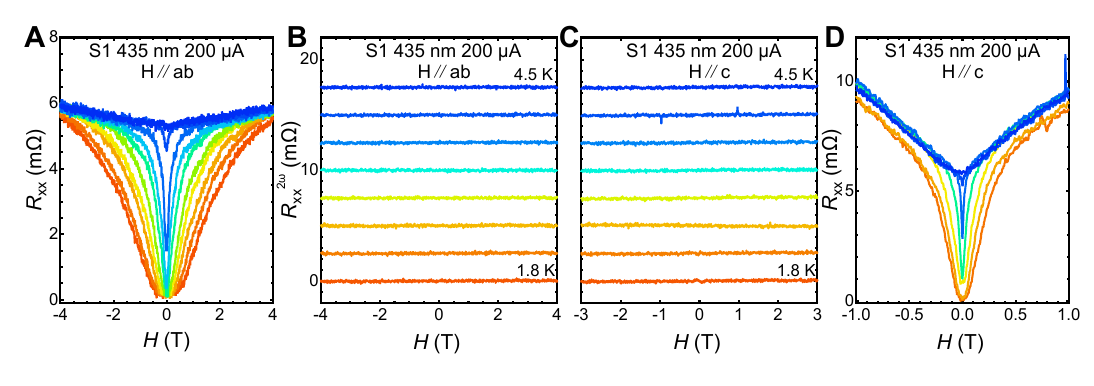}                	\caption{\label{FigS1}\textbf{Absence of nonreciprocity with a higher current density in the 435 nm thin flake.} Resistance and the odd components of the second harmonic signals in the 435 nm \CsVSb\ thin flake with 200 $\mu \rm A$ applied current, confirming the absence of nonreciprocity even with a higher current density.
              }
\end{figure}

\begin{figure}[h]\centering
              \includegraphics[width=1\textwidth]{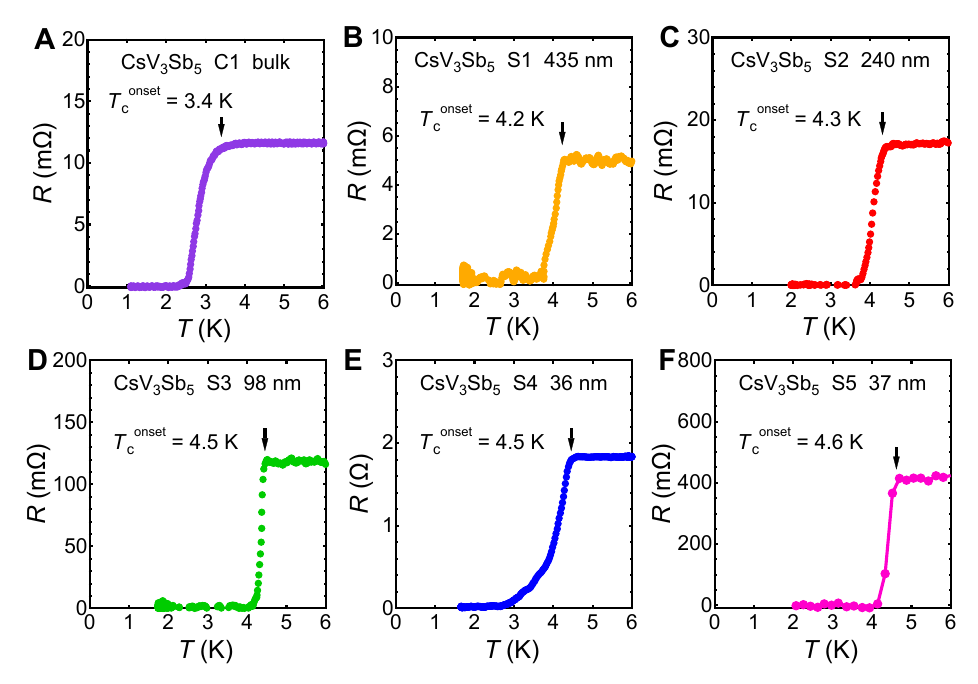}                	\caption{\label{FigS2}\textbf{Temperature dependence of resistance for \CsVSb\ samples.}
              }
\end{figure}

\begin{figure}[h]\centering
              \includegraphics[width=1.03\textwidth]{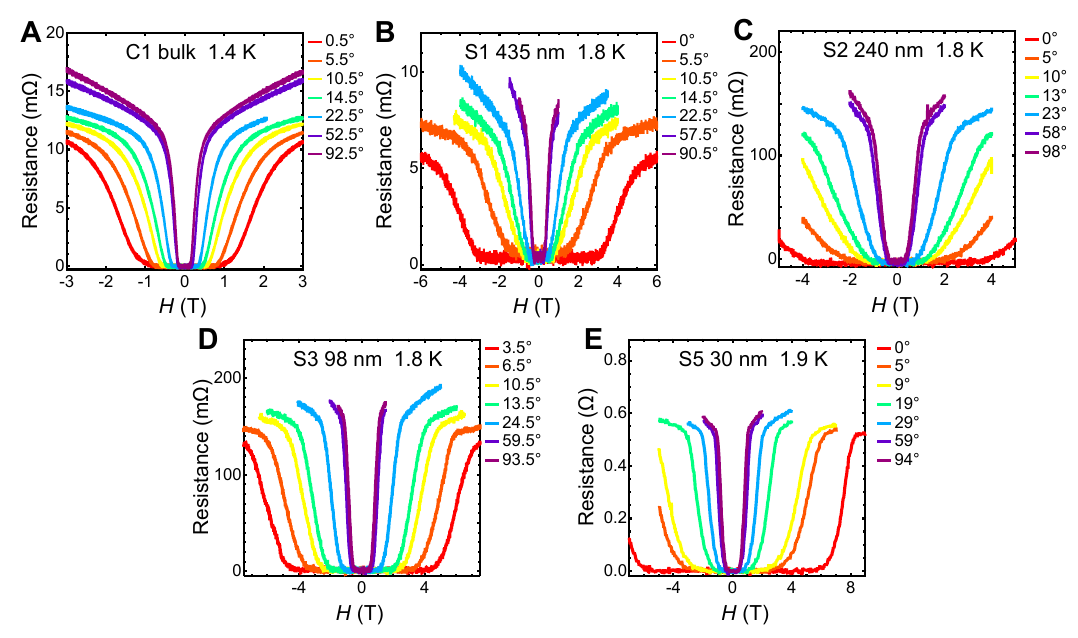}                	\caption{\label{FigS3}\textbf{Field dependence of resistance for different \CsVSb\ samples at representative angles.}
              }
\end{figure}

\begin{figure}[h]\centering
              \includegraphics[width=0.9\textwidth]{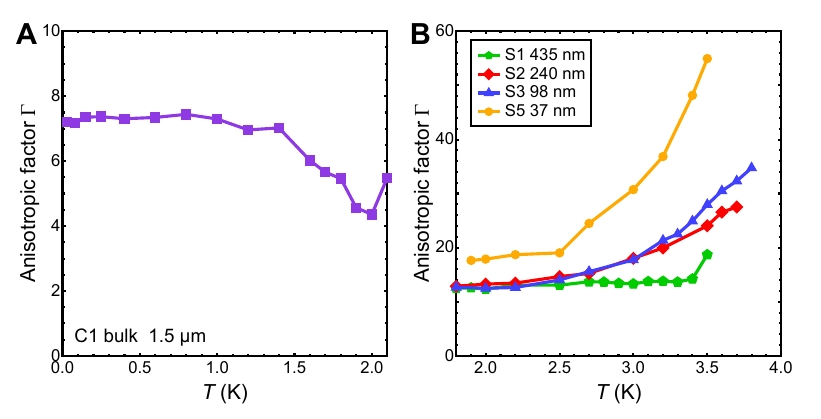}                	\caption{\label{FigS4}\textbf{Temperature dependence of anisotropic factors for bulk and thin flakes.}
              }
\end{figure}

\begin{figure}[h]\centering
              \includegraphics[width=1\textwidth]{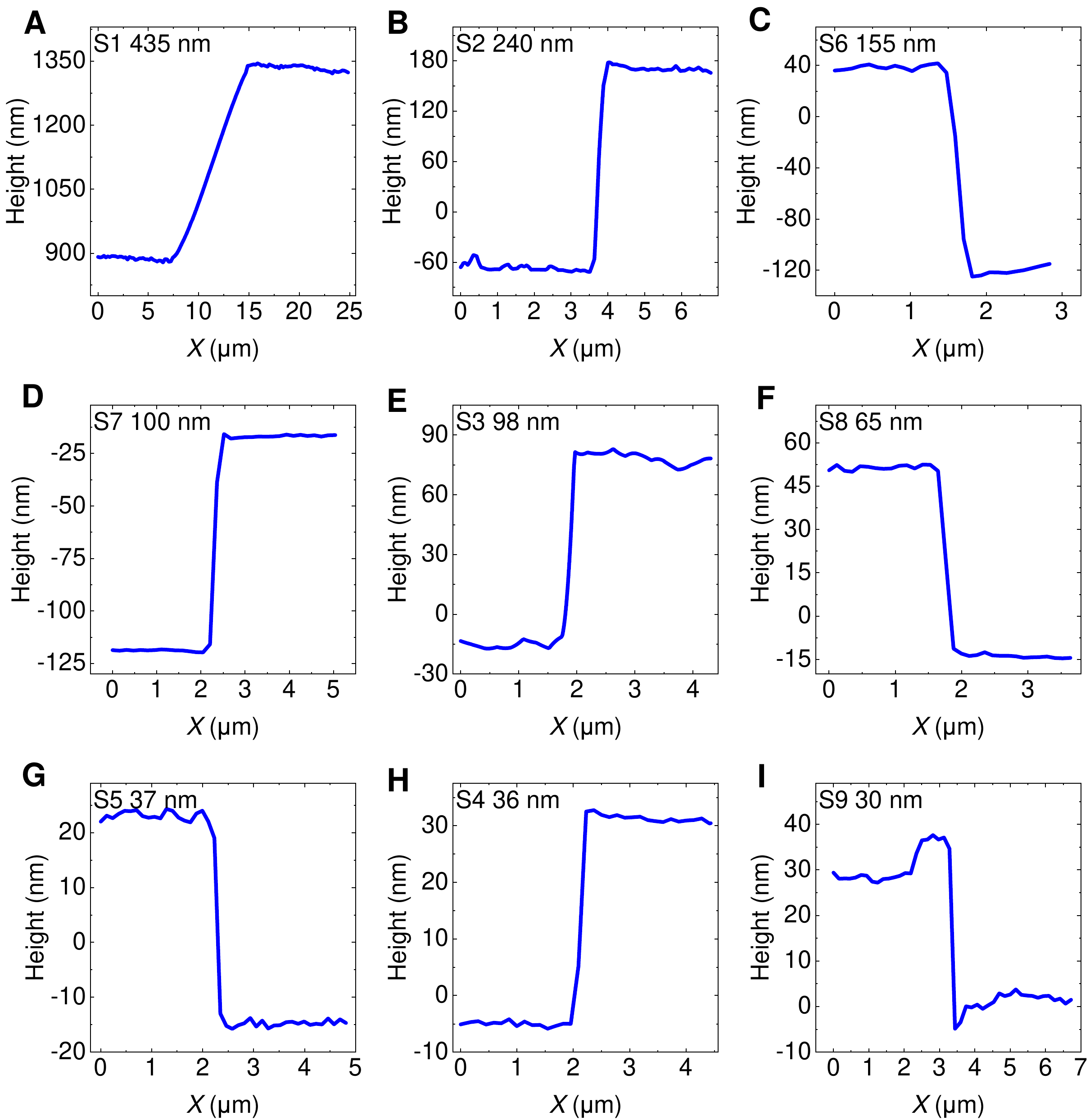}                	\caption{\label{FigS5} \textbf{Thicknesses of \CsVSb\ thin flake devices measured by atomic force microscopy.}
              }
\end{figure}


\clearpage 



\end{document}